\def\kms{\nobreak\mbox{$\;$km\,s$^{-1}$}}
\def\la{\mathrel{\hbox{\rlap{\hbox{\lower4pt\hbox{$\sim$}}}\hbox{$<$}}}}
\def\ga{\mathrel{\hbox{\rlap{\hbox{\lower4pt\hbox{$\sim$}}}\hbox{$>$}}}}
\def\mag{\hbox{$.\!\!^{\rm m}$}}

\documentstyle[11pt,newpasp,psfig,twoside]{article}
\begin{document}

\title{Cepheids, Supernovae, {\boldmath $H_0$}, and the Age of the Universe}
\author{G.\,A. Tammann, B. Reindl, F. Thim}
\affil{Astronomisches Institut der Universit\"at Basel \\
Venusstrasse 7, CH-4102 Binningen, Switzerland}
\author{A. Saha}
\affil{National Optical Astronomy Observatories \\
950 North Cherry Ave., Tucson, AZ 85726}
\author{A. Sandage}
\affil{The Observatories of the Carnegie Institution of Washington \\
813 Santa Barbara Street, Pasadena, CA 91101}

\begin{abstract}
The local expansion field is mapped using Cepheids, a complete sample
of TF distances, and nearby cluster distances. The large-scale field
is mapped using Cepheid-calibrated blue SNe\,Ia. These data give
$H_0({\rm local})=59.2\pm1.4\;[\mbox{km\,s}^{-1}\,\mbox{Mpc}^{-1}]$
and $H_0({\rm cosmic})=57.4\pm2.3$. The intermediate expansion field
($1200\le v \la 10\,000\kms$) is less well calibrated but fully
consistent with $H_0\approx60$. $H_0$ is therefore (nearly)
scale-invariant (high-density regions excluded). 
-- The P-L relation of Cepheids is based on an improved zero point of
$(m-M)_{\rm LMC}=18.56$. 
The slope of the P-L relation for $P>10^{\rm d}$, as judged from OGLE
data (Udalski et~al. 1999) is flatter than anticipated, which tends to
increase the above values of $H_0$ by 3.4 units. No significant
metallicity effect on the Cepheid distances seems to be indicated.
For all practical purposes $H_0=60$ is recommended with a
systematic error of probably less than 10\%. -- The corresponding
expansion age is $T=15.7\pm1.5\;$Gy 
(with $\Omega_{\rm m}=0.3$, $\Omega_{\Lambda}=0.7$),
which compares well with the formation time of $15\pm2\;$Gy for the
apparently oldest globular cluster M\,107. 
\end{abstract}

\section{Introduction}
%
Cepheids and their $P$-$L$ relation are the principal fundament of
extragalactic distances (Section~2). Cepheid distances are available
of nine galaxies which have produced blue SNe\,Ia (Saha et~al\ 2001
and references therein). Their resulting,
very uniform luminosities can be applied to 35 more distant SNe\,Ia
($v \la 30\,000\kms$) to yield a first rate determination of the
large-scale value of $H_0$ (Section~5). Details of the local expansion
field are inserted in Section~3 and some remarks on the not so local
expansion field in Section~4. The expansion age of the Universe is
compared with independent age determination of old objects in the
Galaxy in Section~6. Results and conclusions are compiled in Section~7.

\section{Cepheids and their P-L relation}\label{sec:2}
%
  \subsection{The shape of the P-L relation}\label{sec:2_1}
%
The available Cepheid distances, which are useful for the
extragalactic distance scale beyond the Local Group, are compiled in 
Table~\ref{tab:1}. 
The majority is derived from {\sl HST\/} observations in $V$ and
$I$. All distances are based on the shape of the $P$-$L$ relations by
Madore \& Freedman (1991)
\begin{eqnarray}\label{eq:1}
   M_V & = & -2.76\,\log P - 1.46 \\ \label{eq:2}
   M_I & = & -3.06\,\log P - 1.87 
\end{eqnarray}
As for the constant term see 2.2. Equations~(\ref{eq:1}) and
(\ref{eq:2}) yield apparent distance moduli $\mu_{V}$ and
$\mu_{I}$ which, with $A_{B}=4.1\,E_{B-V}$, $A_{V}=3.1\,E_{B-V}$, 
and $A_{I}=1.8\,E_{B-V}$, give the true moduli
\begin{equation}\label{eq:3}
   \mu^0 \equiv (m-M)^0=2.38\mu_{I} - 1.38\mu_{V}\,.
\end{equation}

It may be noted that equations~(\ref{eq:1}) and (\ref{eq:2}) imply
$(V-I)\propto 0.3\log P$. 
It was recently suggested (Freedman
et~al.\ 2001) that the OGLE photometry of Cepheids in LMC (Udalski
et~al.\ 1999) required a shallower relation $(V-I)\propto 0.2\log P$,
\begin{table}[bht]
\begin{center}
\begin{minipage}{1.05\textwidth}
\scriptsize
\caption{Adopted Cepheid Distance Moduli $\mu^0$[LMC at $\mu^0=18.56$].}
\label{tab:1}

\hspace*{-1cm}
\begin{tabular}{lrccl|lrccl}
\noalign{\smallskip}
\hline\hline
\noalign{\smallskip}
 Galaxy & $v_{220}$ & $\Delta$[O/H]$^{**}$ & $\mu^{0\,\dagger}$ & Src.$^{\ddagger}$ &
 Galaxy & $v_{220}$ & $\Delta$[O/H]$^{**}$ & $\mu^{0\,\dagger}$ & Src.$^{\ddagger}$ \\
\noalign{\smallskip}
\hline
\noalign{\smallskip}
NGC 224   & LG   & $+0.50$ & 24.47 (08) & 5  & NGC 3982  & 1497 & ---   & 31.78 (14) & 1C  \\
NGC 300   & 118  & $-0.15$ & 26.68 (10) & 5  & NGC 4258  & 554  & $+0.35$  & 29.55 (07) & 5   \\
NGC 598   & LG   & $+0.30$ & 24.64 (10) & 5  & NGC 4321  & ---\,\,  & $+0.65$  & 31.02 (07) & 5   \\
NGC 925   & 790  & $+0.05$ & 30.00 (04) & 5  & NGC 4414  & 660  & $+0.70$  & 31.35 (15) & 2C  \\
NGC 1326A & 1338 & $\pm0.00$  & 31.26 (10) & 5  & NGC 4496A & ---\,\,  & $+0.25$ & 31.09 (10) & 1C  \\
NGC 1365  & 1338 & $+0.45$  & 31.42 (05) & 5  & NGC 4527  & ---\,\,  & ---   & 30.73 (15) & 3   \\
NGC 1425  & 1338 & $+0.50$  & 31.73 (06) & 5  & NGC 4535  & ---\,\,  & $+0.70$  & 31.09 (05) & 5   \\
NGC 2090  & 814  & $+0.30$  & 30.50 (04) & 5  & NGC 4536  & ---\,\,  & $+0.35$ & 31.16 (12) & 1C  \\
NGC 2403  & 365  & $+0.30$  & 27.65 (24) & 5  & NGC 4548  & ---\,\,  & $+0.85$ & 31.09 (05) & 5   \\
NGC 2541  & 802  & $\pm0.00$  & 30.48 (07) & 5  & NGC 4603  & 2323 & ---   & 32.67 (11) & 6  \\
NGC 3031  & 141  & $+0.25$  & 27.89 (07) & 5  & NGC 4639  & ---\,\,  & $+0.50$ & 32.09 (22) & 1C   \\
NGC 3198  & 861  & $+0.10$  & 30.90 (08) & 5  & NGC 4725  & ---\,\,  & $+0.40$ & 30.61 (06) & 5   \\
NGC 3319  & 955  & $-0.10$ & 30.85 (09) & 5  & NGC 5253  & 153  & $-0.35$ & 28.06 (07) & 1C  \\
NGC 3351  & 648  & $+0.75$  & 30.04 (10) & 5  & NGC 5457  & 413  & $-0.13$  & 29.36 (10) & 5   \\
NGC 3368  & 802  & $+0.70$  & 30.21 (15) & 4C & NGC 7331  & 1108 & $+0.15$  & 30.96 (09) & 5   \\
NGC 3621  & 495  & $+0.15$  & 29.29 (06) & 5  & IC 4182   & 330  & $-0.10$ & 28.42 (12) & 1C  \\
NGC 3627  & 542  & $+0.75$  & 30.28 (12) & 1C &           &      &       &            &     \\
\noalign{\smallskip}
\hline
\end{tabular}
$^*$ Recession velocity corrected for Virgocentric infall
(Kraan-Korteweg 1986). The velocities of the galaxies in the Virgo cluster
region ($\alpha_{\rm M87}<30^{\circ}$) are not used in the following and are not shown.

$^{**}$ $\Delta\log\mbox{[O/H]}=\log\mbox{[O/H]}_{\rm
  Gal}-\log\mbox{[O/H]}_{\rm LMC}$ derived from data given in Freedman
et~al. (2001) 

$^\dagger$ The errors in parentheses in units of $0\mag01$ are taken
from the original sources. Some of them are unrealistically small. If
the errors of the apparent moduli in $V$ and $I$ are optimistically
assumed to be $\pm0\mag05$, the errors of $(m-M)^0$ are $\ga0\mag15$
from equation~(\ref{fig:3}). -- Consequently the reliable maser
distance $\mu^0=29.29\pm0.08$ of NGC\,4258 (Herrnstein et~al.\ 1999)
is still in statistical agreement with the distance shown here.

$^\ddagger$ C means the galaxy distance is used for the luminosity
calibration of SNe\,Ia.

Sources: (1) Saha et~al.\ 2001; (2) Thim 2001; (3) mean of Saha
et~al.\ 2001 and Thim 2001; (4) mean of Tanvir et~al.\ 1999 and source
(5); (5) Freedman et~al.\ 2001 (Table~3, column~2; including (small)
corrections by excluding short-period Cepheids which may introduce
selection bias); (6) Newman et~al. 1999. 
\end{minipage}
\end{center}
\end{table}
that long-period Cepheids were therefore intrinsicly bluer than
anticipated, and that the corresponding increase of the internal
absorption would lead to a ``dramatic'' decrease of $\sim 0\mag14$ on
average of those Cepheid distances in Table~\ref{tab:1}, which rely
indeed on {\em long\/}-period Cepheids.

   However, the data of Udalski et~al.\ (1999) indicate a clear change
of the $(V-I) - \log P$ relation near $P=10\;$days (Fig.~\ref{fig:1}),
i.e. near the termination point of the Hertzsprung progression. While
the slope is approximately 0.2 for $P<10^{\rm d}$, it steepens to $\ga
0.3$ for $P>10^{\rm d}$. In fact the new data show Cepheids with
$P>10^{\rm d}$ to be {\em redder\/} than implied by
equations~(\ref{eq:1}) and (\ref{eq:2}). (This statement holds if all
Cepheid colors are consistently corrected by a constant LMC excess of
$E_{B-V}=0.1$. The color excesses given by Udalski et~al. are variable
and larger on average. Note that any possible overestimate of
$<\!E_{B-V}\!>$ of the LMC Cepheids affects the apparent moduli
$\mu_{V}$  and $\mu_{I}$ of external galaxies, but {\em not\/} the
true modulus from equation~(\ref{eq:3}). Therefore the Udalski
et~al. values of $E(B-V)$ are adopted in the following without
consequences for the derived distances).
\def\floatwidth{0.68\textwidth}
\begin{figure} 
  \centerline{\psfig{file=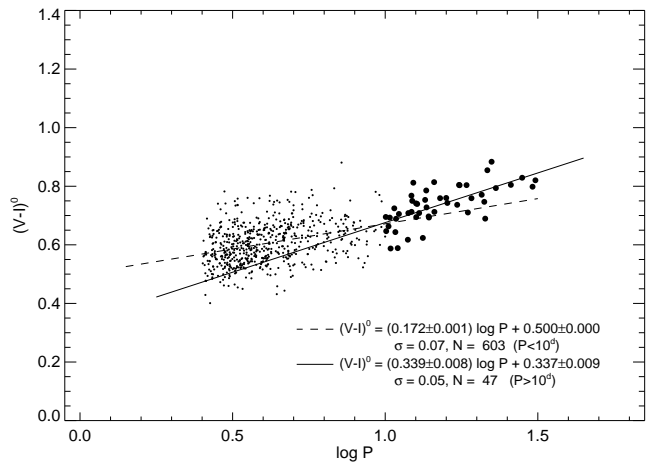,width=\floatwidth}}
\caption{A $(V-I)$ vs.\ $\log P$ diagram of the fundamental-mode LMC
  Cepheids adopted by Udalski et~al.\ (1999). Color excesses from that
  source. Small symbols: $P<10^{\rm d}$; large symbols: $P>10^{\rm d}$.}
\label{fig:1}
\end{figure}

   The change of slope of the color relation in Fig.~\ref{fig:1}
demands a corresponding change of slope near $P=10^{\rm d}$ of the
$M_{V} - \log P$ or of the $M_{I} - \log P$ relation or of
both. Fits to the data for $P<10^{\rm d}$ and $P>10^{\rm d}$ give
indeed significantly different slopes (Fig.~\ref{fig:2}):
\begin{figure} 
  \centerline{\psfig{file=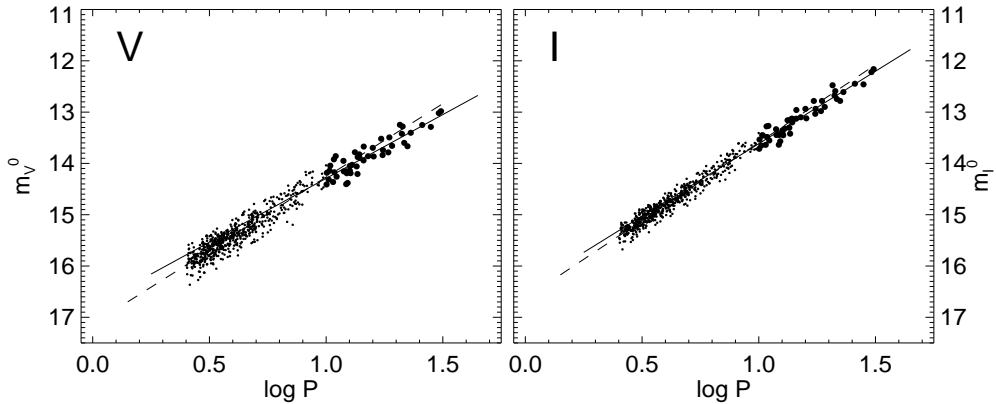,width=1.0\textwidth}}
\caption{$P$-$L$ relations in $V$ and $I$ of the fundamental-mode LMC
  Cepheids adopted by Udalski et~al. (1999). Symbols like in Fig.~1.}
\label{fig:2}
\end{figure}
\begin{eqnarray}
\label{eq:4}
 P<10\;{\rm days}: M_{V}&=&(-2.862\pm0.002)\log P - (1.43\mp0.001); \\
\label{eq:5}
  M_{I}&=&(-3.034\pm0.001)\log P - (1.93\mp0.001) \\ 
\label{eq:6}
 P>10\;{\rm days}: M_{V}&=&(-2.487\pm0.025)\log P - (1.79\mp0.030); \\
\label{eq:7}
  M_{I}&=&(-2.825\pm0.019)\log P - (2.12\mp0.022) 
\end{eqnarray}
Here $\mu^{0}_{\rm LMC}\equiv18.56$ is assumed
(cf.\ 2.2). Compared to equations~(\ref{eq:1}) and
(\ref{eq:2}) the equations~(\ref{eq:6}) and
(\ref{eq:7}) lead to smaller distance moduli by
\begin{equation}\label{eq:8}
   \Delta\mu(P>10^{\rm d}) = -0.18 \log P +0.14.
\end{equation}
This is not because Cepheids are bluer here (they have $(V-I)=1.01$ at
$\log P=2$ in either case even with the large $E_{B-V}$ values of
Udalski et~al.), but because the $P$-$L$ relations are flatter for
$P>10^{\rm d}$ than anticipated.

   In the following the $P$-$L$ relations of equation~(\ref{eq:1}) and
(\ref{eq:2}) are maintained because the data of Udalski et~al.\ (1999)
extend to only $P\le30\;$days, while about half of the Cepheids in
more distant galaxies have longer periods. However, the effect of the
equations~(\ref{eq:6}) and (\ref{eq:7}) -- on the assumption that they
can be extrapolated to longer periods -- is taken up again in
Section~7.

  \subsection{The zero point of the P-L relation}\label{sec:2_2}
%
It is customary to set the zero point of the $P$-$L$ relation by adopting
$\mu_{\rm LMC}=18.50$. It is time now to improve this value. LMC
moduli are compiled in Table~\ref{tab:2} comprising a variety of
methods with different, but poorly known systematic errors. A
unweighted mean of $\mu^{0}_{\rm LMC}=18.56$ is therefore adopted,
which is used throughout this paper.

\begin{table}[thb]
\caption{The Distance Modulus of LMC}
\label{tab:2}
\begin{center}
\scriptsize
\begin{tabular}{lll}
\hline\hline
\noalign{\smallskip}
Reference & \multicolumn{1}{c}{$\mu^0$} & Method \\
\noalign{\smallskip}
\hline
\noalign{\smallskip}
Sandage \& Tammann (1971) & $18.59$   & Galactic Cepheids in Clusters \\
Feast (1984) & $18.50$ & Cepheids, OB, RR\,Lyrae, MS-fitting, Mira \\
Feast \& Catchpole (1997) & $18.70\pm0.10$ & HIPPARCOS parallaxes of
Galactic Cepheids \\
Van Leeuwen et~al.\ (1997) & $18.54\pm0.10$ & HIPPARCOS parallaxes of
Miras \\
Madore \& Freedman (1998) & $18.44-18.57$ & HIPPARCOS parallaxes of
Galactic Cepheids \\
Panagia (1999) & $18.58\pm0.05$ & Ring around SN\,1987A \\
Pont (1999) & $18.58\pm0.05$ & HIPPARCOS parallaxes of Galactic Cepheids \\
Walker (1999) & $18.55\pm0.10$ & Review \\
Feast (1999) & $18.60\pm0.10$ & Review \\
Walker (1992), Udalski et~al.\ (1999) & $18.53\pm0.08$ & $M_{V}^{\rm
  RR}=0.41\pm0.07$ Sandage (1993), \\ 
       & & Chaboyer et~al.\ (1998) \\
Groenewegen \& Oudmaijer (2000) & $18.60\pm0.11$ & HIPPARCOS
parallaxes of Galactic Cepheids \\
Kov{\'a}cs (2000) & $18.52$ & Double mode RR Lyr \\
Sakai et~al.\ (2000a) & $18.59\pm0.09$ & Tip of RGB \\
Cioni et~al.\ (2000) & $18.55\pm0.04$ & Tip of RGB \\
Romaniello et~al. (2000) & $18.59\pm0.09$ & Tip of RGB \\
Groenewegen \& Salaris (2001) & $18.42\pm0.07$ & Eclipsing binary \\
Girardi \& Salaris (2001) & $18.55\pm0.05$ & Red clump stars \\
Baraffe \& Alibert (2001) & $18.60-18.70$ & Pulsation theory of
Cepheids \\
\noalign{\smallskip}
\hline
\noalign{\smallskip}
Adopted & $18.56\pm0.02$ & \\
\noalign{\smallskip}
\hline
\end{tabular}
\end{center}
\end{table}

   Red-clump stars and binaries were suggested to give rather low LMC
moduli. Yet the most recent results agree statistically with the
adopted value (cf.\ Table~\ref{tab:1}). 
   Statistical parallaxes of Galactic RR\,Lyr stars, which pose
formidable problems of sample selection, are given zero weight
(cf.\ Walker 1999).

  \subsection{Metallicity effects of Cepheids?}\label{sec:2_3}
%
The metal dependence of the $P$-$L$ relation has so far not found a
satisfactory solution from observations (e.g.\ Gould 1994; Sasselov
et~al.\ 1997; Kennicutt et~al.\ 1998). Theoretical models give only a
small effect (Chiosi et~al.\ 1993; Saio \& Gautschy 1998; Sandage
et~al.\ 1999; Alibert \& Baraffe 2000; see however Caputo et~al.\
2000). 
If the model dependencies in $V$ and $I$ from Sandage et~al.\ (1999,
Fig.~3) are inserted into equation~(\ref{eq:3}) one obtains a metal
correction of $\Delta \mu = -0.12\Delta$[Fe/H] in formal agreement with
$-0.24\pm0.16\Delta$[O/H] from Kennicutt et~al.\ (1998). The rather
marginal evidence from Fig.~3 has, however, the opposite
sign. Moreover, the seven galaxies with Cepheid distances and known
$\Delta$[O/H] in Table~\ref{tab:1}, for which also SNe\,Ia distances can be
determined based on a {\em mean\/} $M^{\rm corr}$ value
(equation~(\ref{eq:13}) below), give $-0.03\pm0.04\Delta$[O/H] with a
fortuitously small error. Even if the coefficient was as large as
$\pm0.25$ the galaxy distances in Table~\ref{tab:1} would be 
increased/decreased by only $0\mag08$ on average. No metallicity
correction is applied in the following.  

\def\floatwidth{0.75\textwidth}
\begin{figure} 
  \centerline{\psfig{file=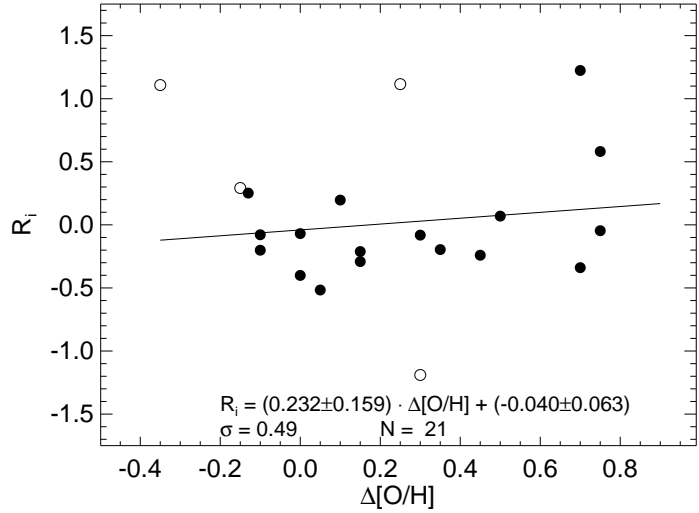,width=\floatwidth}}
\caption{The dependence of Cepheid distances on the metallicity
  $\Delta\mbox{[O/H]}=\mbox{[O/H]}_{\rm Gal}-\mbox{[O/H]}_{\rm
    LMC}$. 
  Plotted are the residuals $R_{\rm i}$ from the mean
  Hubble line in Fig.~4. Open symbols for galaxies with $\mu^0<28.1$
    are given half weight.} 
\label{fig:3}
\end{figure}

\section{The Local Expansion Field {\boldmath $(v_{220}\la1200\kms)$}}
\label{sec:3}
%
  \subsection{The local distance-calibrated Hubble diagram of Galaxies
              with Cepheid distances}\label{sec:3_1} 
\begin{figure} 
  \centerline{\psfig{file=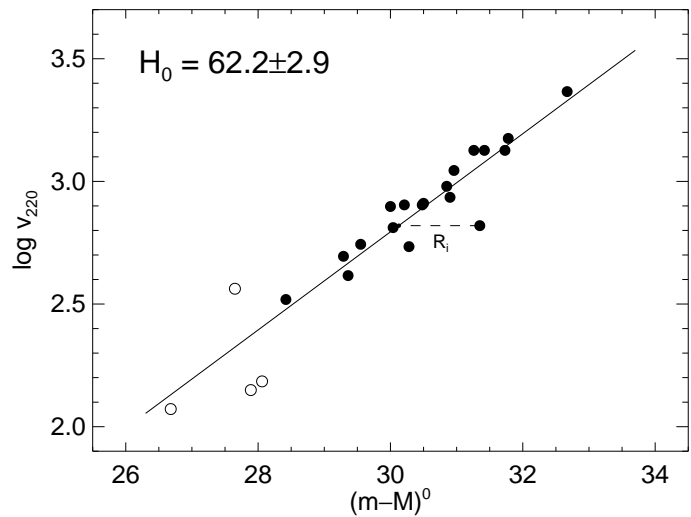,width=\floatwidth}}
\caption{The distance-calibrated Hubble diagram plotting $\log
  v_{220}$ versus the Cepheid distance modulus of 23 galaxies. Open
  symbols are given half weight. The distance modulus residuals
  $R_{\rm i}$ are used in Fig.~3.} 
\label{fig:4}
\end{figure}

%
The values $\log v_{220}$ of the galaxies in Table~\ref{tab:1} are
plotted versus their Cepheid distances $\mu^0$ in
Fig.~\ref{fig:4}. Galaxies within $30^{\circ}$ of the Virgo cluster
center are omitted because that region is noisy due to virial motions
and infall from the cluster front- and back-sides.

   If a Hubble line of slope 0.2 is fitted to the 23 galaxies with
Cepheid distances, giving half weight to galaxies with $(m-M)^0<28.1$,
one obtains
\begin{equation}\label{eq:9}
  \log v = 0.2 (m-M)^0 + a_0
\end{equation}
with $a_0=-3.206\pm0.020$. Since $\log H_0=a_0 +5$, the local value of
$H_0$ becomes $H_0=62.2\pm2.9\; [\mbox{km\,s}^{-1}\,\mbox{Mpc}^{-1}]$.

   The scatter about the Hubble line amounts to
$\sigma_{\mu}=0\mag43$, of which $\sigma_{\mu}=0\mag15-0\mag20$
can be attributed to errors of the Cepheid moduli (cf. footnote in
Table~\ref{tab:1}). The contribution to the scatter by peculiar
velocities becomes then $\sigma_{\log v}\la0.08$ or $\Delta v_{\rm
pec}/v_{220}\la20\%$. 

   If the velocities $v_0$ (corrected only to the barycenter of the
Local Group) were used instead of $v_{220}$ one would obtain
$H_0=59.0\pm2.9$ with slightly larger scatter. 

  \subsection{The local distance-calibrated Hubble diagram from the TF
    relation} 
  \label{sec:3_2}
%
The Tully-Fisher (TF) relation ($M_{\rm B}^0$ vs. 21-cm line width at
20\% maximum intensity) can be calibrated via 29 sufficiently
inclined spiral galaxies from Table~\ref{tab:1}. Two galaxies,
NGC\,5204 and 5585, can be added with half weight on the assumption
that they lie as companions of NGC\,5457 at the latter's distance. One
obtains then
\begin{equation}\label{eq:10}
  M_{\rm B}^0=-7.31 \log w_{20} - (1.833\pm0.095); \quad \sigma_{\rm M}=0.53.
\end{equation}
The data of the individual galaxies are given by Federspiel
(1999). The slope of the relation is taken from a complete sample of
Virgo cluster members (Federspiel 1999) and is in very good agreement
with the calibrators used here. 
\def\floatwidth{1.0\textwidth}
\begin{figure} 
  \centerline{\psfig{file=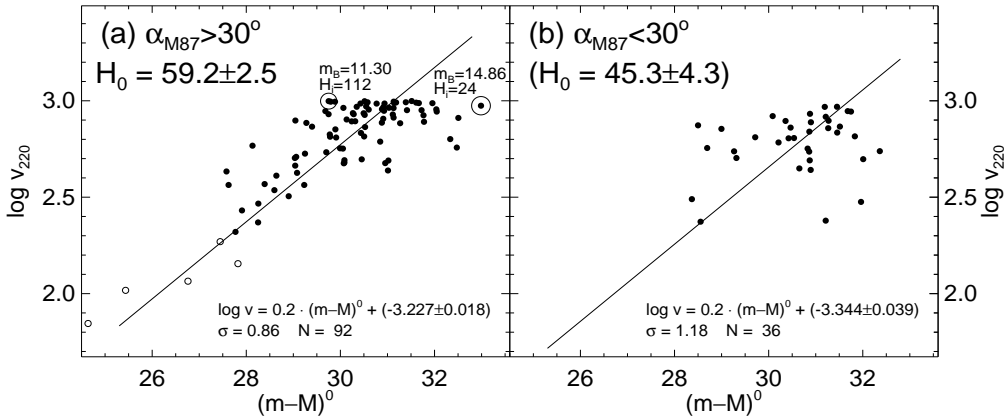,width=\floatwidth}}
\caption{The distance-calibrated Hubble diagram of a {\em complete\/}
  sample of spiral galaxies with TF distances and with
  $v_{220}<1000\kms$. (a) Galaxies outside the Virgo region
  ($\alpha_{\rm M87}>30^{\circ}$). An apparently bright and an
  apparently faint galaxy and their respective individual Hubble
  constants $H_{\rm i}$ are shown. (b) Galaxies within the Virgo
  region ($\alpha_{\rm M87}<30^{\circ}$).}  
\label{fig:5}
\end{figure}

   The calibration in equation~(\ref{eq:10}) is applied to a (nearly)
{\em complete}, distance-limited sample of inclined ($i\ge45^{\circ}$)
spirals with $v_{220}<1000\kms$ and $|b|>20^{\circ}$, as compiled
by Federspiel (1999); Virgo and UMa cluster members are excluded as
well as peculiar and HI-truncated galaxies.
The resulting Hubble diagram of $\log v_{220}$ versus the TF distances
of the 92 galaxies {\em outside\/} the Virgo region ($\alpha_{\rm M87}
> 30^{\circ}$) gives a mean value of $H_0=59.2\pm2.5$
(Fig.~\ref{fig:5}a); the large scatter of $\sigma_{(m-M)}=0.86$ is
presumably due to observational errors of the linewidths $w_{20}$,
which are compiled from various sources. Yet the scatter of the 36
galaxies {\em within\/} $\alpha_{\rm M87}<30^{\circ}$ is still
significantly larger ($\sigma_{(m-M)}=1.18$; Fig.~\ref{fig:5}b).
This can only be explained by the enhanced influence of peculiar
motions in this dense region. In that case the $1000\kms$ limit
provides a fuzzy distance limit and the corresponding value of
$H_0=45.3\pm4.3$ is unreliable.

\begin{figure} 
  \centerline{\psfig{file=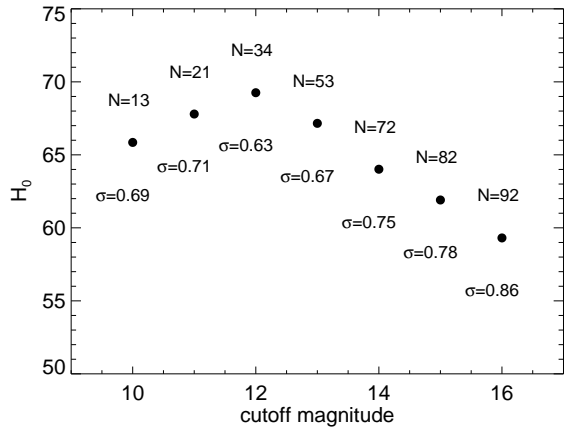,width=0.62\textwidth}}
\caption{The local value of $H_0$ from TF distances 
  ($\alpha_{\rm M87}>30^{\circ}$) in function of the 
  cutoff magnitude of the galaxy sample. Incomplete samples give
  arbitrarily large values of $H_0$.} 
\label{fig:6}
\end{figure}

   The large scatter of the TF distances make the use of {\em
complete}, distance-limited samples essential (cf. Fig.~\ref{fig:6}).

   The statistical agreement of the local value of $H_0$ from Cepheids
(Fig.~\ref{fig:4}) and the TF method (Fig.~\ref{fig:5}) is a
tautology, -- at least to the extent that the fewer Cepheids represent
a fair sample with respect to the local expansion field. The bulk of
the calibrators ($n=31$) of the TF relation have also been used in
Fig.~\ref{fig:4} ($n=23$), and the TF distances can hardly provide
more than a consistency check. 

   Thus Section~3.2 sheds more light on the TF {\em
method\/} than on the local value of $H_0$. It is questionable whether
other distance indicators (planetary nebulae, the $D_{\rm n}-\sigma$
or fundamental plane method, surface brightness fluctuations) will
ever do better for field galaxies -- even if objectively selected
samples will become available -- than the TF method whose
difficulties are apparent here.

  \subsection{The distance-calibrated Hubble diagram of local groups
    and clusters} 
  \label{sec:3_3}
\begin{figure} 
  \centerline{\psfig{file=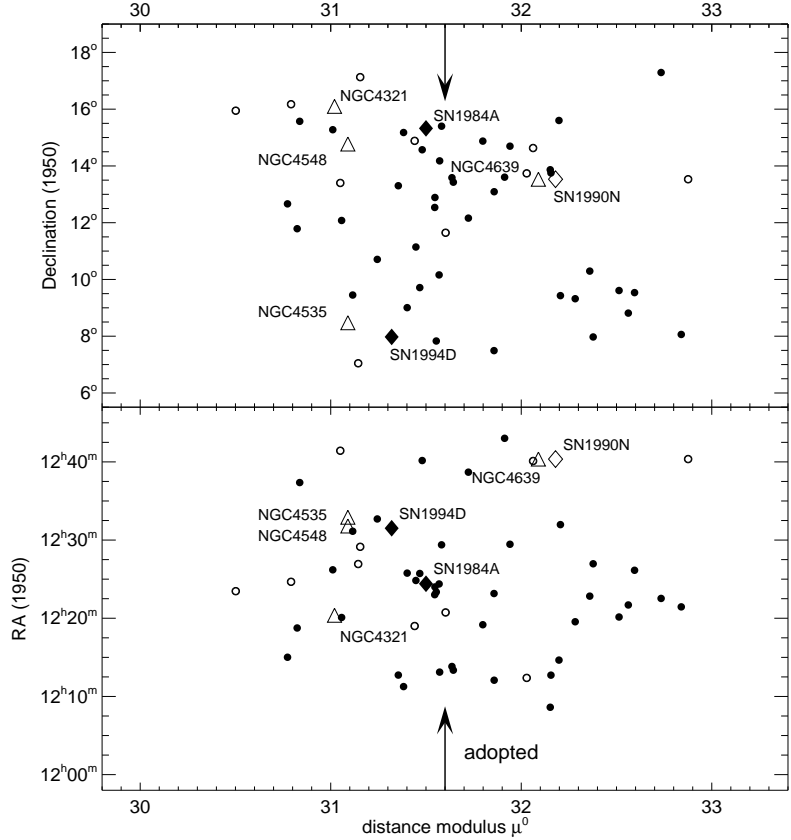,width=0.80\textwidth}}
\caption{An illustration of the depth effect of the
  Virgo cluster. The position of cluster members is shown (declination
  [upper panel] and RA [lower panel] vs.\
  $\mu^0$) with known Cepheid (triangles), SNe\,Ia (diamonds), or
  TF (circles) distances. The majority of the well resolved
  galaxies (open symbols) lie on the near (left) side of the cluster.} 
\label{fig:7}
\end{figure}

%
Useful distance determinations of six local groups and clusters are
compiled in Table~\ref{tab:3}. The Table is self-explanatory. However,
one remark is in place concerning the Virgo cluster. The cluster has
an important depth effect which has caused the {\em mean\/} cluster
distance (i.e., the mean of subclusters A\,$+$\,B) to be controversial
for a long time.  
The well resolved cluster spirals (Sandage \& Bedke 1988) are expected
to lie on the near side of the cluster; this expectation is born out
for the majority of spirals with known Cepheid, SNe\,Ia or TF
distances (Fig.~\ref{fig:7}). The Cepheid distances of NGC\,4321,
4535, and 4548, which were selected on grounds of their high
resolution can therefore not reflect the {\em mean\/} cluster distance.

\begin{table}
\caption{Distances of local groups and clusters}
\label{tab:3}
\begin{center}
\begin{minipage}{0.95\textwidth}
\scriptsize
\begin{tabular}{lrrl}
\hline\hline
\noalign{\smallskip}
Group/Cluster & $<\!v_{220}\!>$ & \multicolumn{1}{c}{$\qquad\mu^0$} & Method \\
\noalign{\smallskip}
\hline
\noalign{\smallskip}
South Polar gr.$^*$ &  $112$ & $26.68\pm0.20$    & Cepheids (1)$^{\rm a}$ \\
M\,101 gr.$^*$      &  $405$ & $29.36\pm0.10$    & Cepheids (1)$^{\rm a}$,
                                                   TRGB$^{\rm f}$ \\
Leo$^*$             &  $652$ & $30.21\pm0.05$    & SNe\,Ia (2)$^{\rm b}$, 
                             Cepheids (3)$^{\rm a}$, TRGB$^{\rm f}$ \\
UMa$^{\dagger}$     & $1060$ & $\ge31.33\pm0.15$ & TF$^{\rm c}$, 
                                                   LC$^{\rm d}$ \\
Virgo$^{\ddagger}$  & $1179$ & $31.60\pm0.20$    & SNe\,Ia (3)$^{\rm b}$,
                                    Cepheids (4)$^{\rm a}$, TF$^{\rm
                                      c}$, LC$^{\rm d}$, GC$^{\rm e}$ \\
Fornax$^*$          & $1338$ & $31.60\pm0.10$    & SNe\,Ia (3)$^{\rm b}$, 
                                                   Cepheids (3)$^{\rm a}$\\ 
\noalign{\smallskip}
\hline
\noalign{\smallskip}
\end{tabular}

$^*$ Group/cluster membership as defined by Kraan-Korteweg (1986)

$^{\dagger}$ Two separate agglomerations can be distinguished in the
UMa cluster; the nearer, presumably more nearly complete agglomeration
with 18 TF distances is considered here (Federspiel 1999)

$^{\ddagger}$ Cluster membership as defined by Binggeli et~al. (1993)

Sources: (a) Table~1; (b) SNe\,Ia calibration as in Section~5.4; (c)
Federspiel (1999); (d) Luminosity classes (Sandage 2001); (e) Globular
Clusters (Tammann \& Sandage 1999); (f) Kennicutt et~al.\ (1998)
\end{minipage}
\end{center}
\end{table}

\def\floatwidth{0.70\textwidth}

   The Virgo cluster distance from the peak of the luminosity function
of globular clusters $(\mu^0=31.70\pm0.30)$ is of particular interest
because it is here the only distance independent of Cepheids, being
based on only Population~II objects (RR\,Lyr and Galactic globular
clusters). Yet the method still lacks a theoretical understanding and
does not seem to work for the Fornax cluster (Tammann \& Sandage 1999).

\begin{figure}
  \centerline{\psfig{file=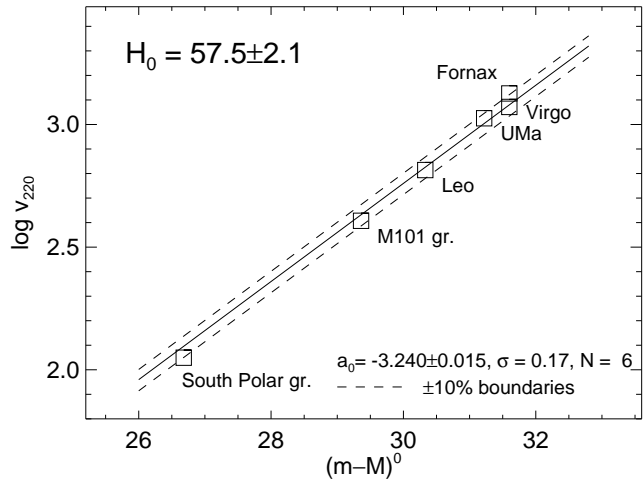,width=\floatwidth}}
\caption{The distance-calibrated Hubble diagram for six local groups and
  clusters from Table~3.} 
\label{fig:8}
\end{figure}

   The Hubble diagram of the six groups and clusters in Table~\ref{tab:3}
is shown in Fig.~\ref{fig:8}. Giving half weight to the nearby South
Polar group (NGC\,300) yields $H_0=57.5\pm2.1$.

   The three determinations of the local value of $H_0$ in
3.1 to 3.3 give a weighted mean of 
$H_0=59.2\pm1.4$.

\section{The Intermediate Expansion Field} 
  \label{sec:4}
%
For the determination of the {\em cosmic\/} value of $H_0$ the
intermediate expansion field ($1200\la v_{220}\la10\,000\kms$) is
irrelevant. The large-scale route through SNe\,Ia is in any case
superior and the most direct one (Section~5). Intermediate-size
distances are only needed to study deviation from a smooth Hubble flow
(e.g.\ the local CMB dipole motion). SNe\,Ia are not yet useful here
because of their paucity.

   Distance determinations of field galaxies beyond $\sim\!1200\kms$
are extremely difficult because of incompleteness bias. Elaborate
parametric or non-parametric methods have been developed to correct
for this bias (e.g.\ Theureau et~al.\ 1998; Federspiel et~al.\ 1998;
for a compilation cf.\ Tammann et~al.\ 2000; Hendry 2001). The
conclusion is $50<H_0<65$.

   Nevertheless usefull Hubble diagrams of clusters have been
established using {\em relative\/} cluster distances, which can be
fitted to nearby clusters with known distances (cf.\
Table~\ref{tab:3}) to yield distance-calibrated Hubble diagrams. 
(1) The Hubble diagram of first-ranked cluster galaxies by Sandage \&
Hardy (1973), if fitted to the brightest galaxies in the Virgo, Fornax
(and Coma) clusters, gives $H_0=51\pm7$.
(2) Giovanelli et~al.\ (1997) and Dale et~al.\ (1999) have provided TF
distances of 71 clusters ($v<25\,000\kms$), using about 10 galaxies
per cluster. These clusters define an impressively tight Hubble
diagram with a scatter of only $\sigma_{(m-M)}=0.11$. Using the Fornax
cluster as zero point one obtains $H_0=63\pm5$. (3) Relative D$_{\rm
  n}-\sigma$ and fundamental plane distances have been derived for 10
clusters ($v<10\,000\kms$) by Kelson et~al.\ (2000). The ensuing
Hubble diagram with $\sigma_{(m-M)}=0.19$ combined with the Coma
cluster distance gives $H_0=66\pm8$. For details see Tammann (2001).

   The routes through (1) and (2) go beyond what we have defined as the
``intermediate'' expansion field and reflect already on the
large-scale value of $H_0$. It has been suggested (Sakai et~al.\ 2000b)
that method (2), if fitted to equation~(\ref{eq:10}), leads to
$H_0\ga70$. 
This is, however, a grave error, because equation~(\ref{eq:10}) can
only be applied to distance- or volume-limited samples, while the
relative TF distances of distant clusters are based on only a few
subjectively selected cluster members.  

\section{The Cosmic Expansion Field} 
  \label{sec:5}
%
  \subsection{The Hubble diagram of blue SNe\,Ia out to {\boldmath
      $30\,000\kms$}}  
  \label{sec:5_1}
%
A sample of 35 blue SNe\,Ia with good $B$, $V$ (and $I$) photometry,
$1200\le v_{220}\la 30\,000\kms$, and $(B-V)^0_{\max}\le0.06$
has been compiled by Parodi et~al.\ (2000). Their residual magnitudes
(at maximum light) from a mean Hubble line correlate with the decline
rate $\Delta m_{15}$ and the color $(B-V)^0_{\max}$. If these
dependencies are removed one obtains corrected magnitudes $m^{\rm
  corr}_{B,V,I}$, which define linear Hubble lines (corresponding
closely to an $\Omega_{\rm m}=1$ model) of the form
\begin{equation}\label{eq:11}
   \log v = 0.2\,m^{\rm corr}_{B,V,I} + c_{B,V,I}
\end{equation}
with very small scatter $\sigma_{B,V}=0\mag13$, $\sigma_{I}=0\mag12$
(Parodi et~al.\ 2000). 

   The small scatter is most remarkable because it can be accounted
for by photometric errors alone, leaving very little room for
peculiar motions ($v_{\rm pec}/V_{220}<0.06$) and luminosity
variations of SNe\,Ia, once they are corrected for $\Delta m_{15}$
and color. Yet the most important fact for the determination of $H_0$
is that the small scatter provides indisputable proof for blue SNe\,Ia
(corrected for $\Delta m_{15}$ and color) being by far the best
distance indicators known, which at the same time makes them
insensitive to selection bias.

   Nine of the 35 SNe\,Ia are corrected for Galactic absorption by
more than $A_{V}=0\mag2$ (Schlegel at~al. 1998). They lie
significantly ($0\mag21\pm0.04$) above the mean Hubble line defined by
the remaining 26 SNe\,Ia, presumably because their Galactic absorption
is overestimated. The more reliable 26 SNe\,Ia with small Galactic
absorption yield the following constant terms in
equation~(\ref{eq:11})
\begin{equation}\label{eq:12}
   c_{B}=0.662\pm0.005, \quad c_{V}=0.661\pm0.005, \quad  c_{I}=0.604\pm0.005
\end{equation}
with a reduced scatter of $\sigma_{\rm m}=0\mag11$(!) in all three
colors.

  \subsection{The luminosity calibration of SNe\,Ia} 
  \label{sec:5_2}
%
For nine local SNe\,Ia Cepheid distances are available (Saha
et~al.\ 2001). After standard corrections (Parodi et~al.\ 2000) for
Galactic and internal absorption and for decline rate $\Delta m_{15}$
and color $(B-V)$, they have the same intrinsic colors
$(B-V)^0_{\max}$ as the distant SNe\,Ia. Their Cepheid-calibrated
absolute magnitudes at maximum become (Saha et~al.\ 2001):
\begin{equation}\label{eq:13}
   <\!\!M_{B}^{\rm corr}\!\!>=-19.56\pm.07, \;
   <\!\!M_{V}^{\rm corr}\!\!>=-19.53\pm.06, \;
   <\!\!M_{I}^{\rm corr}\!\!>=-19.25\pm.09.
\end{equation}

  \subsection{The cosmic value of {\boldmath $H_0$}} 
  \label{sec:5_3}
%
Transforming equation~(\ref{eq:11}) gives
\begin{equation}\label{eq:14}
   \log H_0 = 0.2\,M_{B,V,I} + c_{B,V,I} + 5
\end{equation}
Inserting $c$ from equation~(\ref{eq:12}) and $M$ from
equation~(\ref{eq:13}) yields $H_0(B)=56.2\pm2.5$,
$H_0(V)=56.9\pm2.3$, and $H_0(I)=56.8\pm3.1$.

   If instead of a linear fit ($\Omega\approx1$) the distant SNe\,Ia
are fitted by an $\Omega_{\rm m}=0.3$, $\Omega_{\Lambda}=0.7$ model
(Carroll et~al.\ 1992), the value of $H_0$ is marginally increased
by 0.8 units, such that
\begin{equation}\label{eq:15}
   <\!\!H_0(B,V,I)\!\!>_{\rm cosmic}=57.4\pm2.3.
\end{equation}

  \subsection{Other evidence for the large-scale value of {\boldmath
      $H_0$}}  
  \label{sec:5_4}
%
Models of blue SNe\,Ia predict $M_{B}(\max)=-19.5(\pm0.2)$ (H{\"o}flich
\& Khokhlov 1996; Branch 1998), which is in fortuitous agreement
with the observed value in equation~(\ref{eq:13}) which in turn leads
to the value of $H_0$ in equation~(\ref{eq:15}).

Distances have been derived for an impressive number of clusters from
the Sunyaev-Zeldovich effect. Several reviews agree that the present
data suggest $H_0\approx60$ (Mason et~al.\ 2001; Rephaeli 2001;
Carlstrom 2001).

The solution of $H_0$ from a single lensed quasar is degenerate
as to the distance and lensing mass. Yet a combination of several such
quasars restrict the solution to a rather narrow window near
$H_0\approx55$.

   The fluctuation spectrum of the CMB does not yet give stringent
limits on $H_0$, but values near $H_0=60$ are entirely consistent with
multi-parameter solutions (Netterfield et~al. 2001; Pryke 2001).

\section{The Age of the Oldest Galactic Objects}  
\label{sec:6}
%
The age test requires the oldest objects in the Universe to be younger
than the expansion age. The latter becomes, with $H_0=60\pm5$,
$\Omega_{\rm m}=0.3$, and $\Omega_{\Lambda}=0.7$,
$T=15.7\pm1.5\;$Gy. Various age determinations of very old objects in
the Galaxy are available, e.g.\ Th or U ages of metal-poor stars 
(e.g.\ Cowan et~al.\ 1999; Westin et~al.\ 2000) or
cooling times of White Dwarfs (for a compilation cf. Tammann
2001). The most stringent requirement comes probably from globular
clusters. Sandage (1993) and Chaboyer et~al.\ (2000) give $14.1\pm1.5$
and $11.7\pm1.6\;$Gy, respectively, for several globular clusters
(cf.\ also Gratton et~al.\ 1997; Heasley et~al.\ 2000; Sneden et~al.\
2000), but note that M\,107 seems to be older. If $14\pm2\;$Gy is
adopted for the latter, and if $\sim\!1\;$Gy is added for its
gestation time, a cosmic age of $15\pm2\;$Gy is required. This
requirement is well met by the above expansion age.

\def\floatwidth{0.88\textwidth}
\begin{figure}
  \centerline{\psfig{file=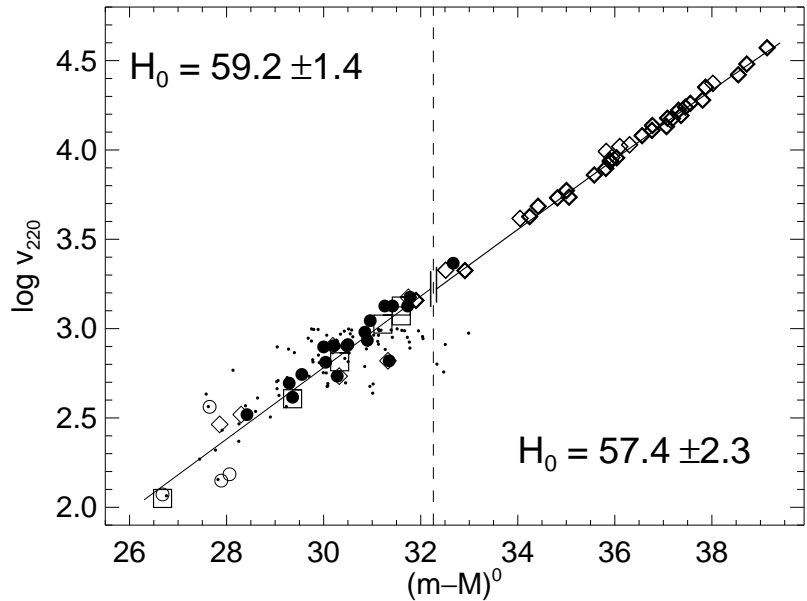,width=\floatwidth}}
\caption{A synoptic distance-calibrated Hubble diagram extending over
  11 magnitudes or a factor of 100 in recession velocity. The
  left-hand value of $H_0$ based on Cepheids (large dots), TF distances
  (small dots), and groups/cluster (squares) is independent of the
  right-hand value of $H_0$ based on SNe\,Ia (diamonds), except that
  all distances rest ultimately on Cepheids. The full drawn line
  corresponds to $\Omega_{\rm m}=0.3$, $\Omega_{\Lambda}=0.7$.} 
\label{fig:9}
\end{figure}
\section{Results and Conclusions}  
\label{sec:7}
%
A synopsis of the distance determinations in Sections~3 and 5 is shown
in Fig.~\ref{fig:9}. The distance moduli of the SNe\,Ia are determined
here from their apparent magnitudes $m^{\rm corr}_{B,V,I}$ and the
absolute magnitudes $M^{\rm corr}_{B,V,I}$ in
equation~(\ref{eq:13}). The local expansion field extends into the
cosmic expansion without any significant break.

   If one choses to base the Cepheid distances of the
SNe\,Ia-calibrating galaxies on the new $P$-$L$ relations
(eq.~\ref{eq:7} \& \ref{eq:8}), they are reduced by 6\% (cf.\
eq.~\ref{eq:8}), i.e.  
\begin{equation}\label{eq:16}
   H_0({\rm cosmic})=60.8\pm2.3.
\end{equation}
For all practical applications $H_0=60$ can be used everywhere, except
in nearby high-density regions.

   So far only statistical errors have been quoted. It comes as a
surprise that the largest source of systematic errors is in the {\em
  shape\/} of the $P$-$L$ relation (6\%), followed by the metallicity
dependence of Cepheids and the photometric HST zero point in the
crowded fields of SNe\,Ia-calibrating galaxies (4\%). The zero point
of the $P$-$L$ relation, 
the slope of the $\Delta m_{15}$ correction of SNe\,Ia and the HST
photometry may each contribute systematic 2-3\% errors. Systematic
errors due to absorption corrections for the nearby, calibrating
SNe\,Ia and the distant SNe\,Ia are negligible, because the two sets
have closely the same colors ($<\!B-V\!>=-0.01\pm0.01$; cf.\ Parodi
et~al.\ 2000). Unless there is a conspiracy of the individual
systematic errors, the total systematic error is $<10\%$.

   The resulting expansion age of $T=15.7\pm1.5\;$Gy ($H_0=60\pm5$,
$\Omega_{\rm m}=0.3$, $\Omega_{\Lambda}=0.7$) gives sufficient room
for the oldest dated objects in the Galaxy.

\smallskip\noindent
{\bf Acknowledgments.}
The first three authors thank the Swiss National Science Foundation
for financial support. G.\,A.\,T. and F.\,T. thank also the PRODEX
programme of the Swiss Space Office for support.


\end{document}